\documentclass[aps, pra, amsmath, amssymb, showpacs, superscriptaddress, reprint]{revtex4-1}

\usepackage{graphicx}
\usepackage{natbib}

\begin{document}

\title{Optical pumping, decay rates and light shifts of cold-atom dark states}
\author{M. Shuker}
\email{moshe.shuker@gmail.com}
\affiliation{National Institute of Standards and Technology, Boulder, Colorado 80305, USA}
\affiliation{University of Colorado, Boulder, Colorado 80309-0440, USA}

\author{J. W. Pollock}
\affiliation{National Institute of Standards and Technology, Boulder, Colorado 80305, USA}
\affiliation{University of Colorado, Boulder, Colorado 80309-0440, USA}

\author{V. I. Yudin}
\affiliation{Novosibirsk State University, ul. Pirogova 2, Novosibirsk, 630090, Russia }
\affiliation{Institute of Laser Physics SB RAS, pr. Akademika Lavrent'eva 13/3, Novosibirsk, 630090, Russia }
\affiliation{Novosibirsk State Technical University, pr. Karla Marksa 20, Novosibirsk, 630073, Russia }

\author{J. Kitching}
\affiliation{National Institute of Standards and Technology, Boulder, Colorado 80305, USA}
\affiliation{University of Colorado, Boulder, Colorado 80309-0440, USA}

\author{E. A. Donley}
\affiliation{National Institute of Standards and Technology, Boulder, Colorado 80305, USA}
\affiliation{University of Colorado, Boulder, Colorado 80309-0440, USA}

\date{\today}

\begin{abstract}
Coherent dark states in atoms, created by simultaneous interaction of two coherent light fields with a 3-level system, are of prime importance in quantum state manipulation. They are used extensively in quantum sensing and quantum information applications to build atomic clocks, magnetometers, atomic interferometers and more. Here we study the formation and decay of coherent dark-states in an ensemble of laser-cooled free-falling atoms. We measure the optical-pumping rate into the dark-state in the $\sigma ^+-\sigma ^-$ polarization configuration. We find that the pumping rate is linear with the optical field intensity, but about an order-of-magnitude slower than the rate predicted by the commonly used, but simplistic, three-level analytic formula. Using a numerical model we demonstrate that this discrepancy is due to the multi-level Zeeman manifold. Taking into account the slower pumping rate we explain quantitatively the relation between the light-shift and the duration of pumping into dark-state in Ramsey spectroscopy. We also measure the decay of the dark-state coherence and find that in our apparatus it is dominated by the mechanical motion of the atoms out of the probing region, while the atomic decoherence is negligible. 
\end{abstract}
\maketitle
\section{Introduction}
\label{Introduction}
Coherent population trapping (CPT) \cite{Arimondo1996} occurs when two resonant light fields interact coherently with a three-level atom, generating a coherent superposition of the ground-state sub-levels (dark state), uncoupled from the light-fields. This phenomenon exhibits ultra-narrow transmission resonances with low peak absorption and enables numerous applications including slow light propagation \cite{HauNature1999}, storage of light in atoms \cite{LukinPRL2001}, miniature microwave atomic clocks \cite{KnappeAPL2004,VanierAPB2005,EsnaultPRA2013}, atom interferometers \cite{KasevichPRL1991} and even unique laser cooling techniques \cite{KulinPRL1997}. In many cases the performance of these measurements is limited by the optical pumping rate and the decay rate of the dark-state superpositions created in the atoms. Therefore it is of prime importance to understand these rates and the physical processes that control them.\\ 
In this study, we use a cold-atom apparatus and measure the pumping and decay rates of dark-states, as well as the associated light shifts, within the ground-state hyperfine levels of $^{87}$Rb. By employing both time-domain and spectroscopic measurements, the underlying processes of the dark-state decay are identified. These measurements were performed on free-falling atoms in vacuum, nearly free from interactions with background gases and other rubidium atoms, which allows a simple comparison to analytical expressions and detailed comparison between decay rates and the CPT line-width.\\ 
\section{Experimental setup and methods}
\label{Experimental setup and methods}
\begin{figure}[t]
    \includegraphics[width=8.6cm]{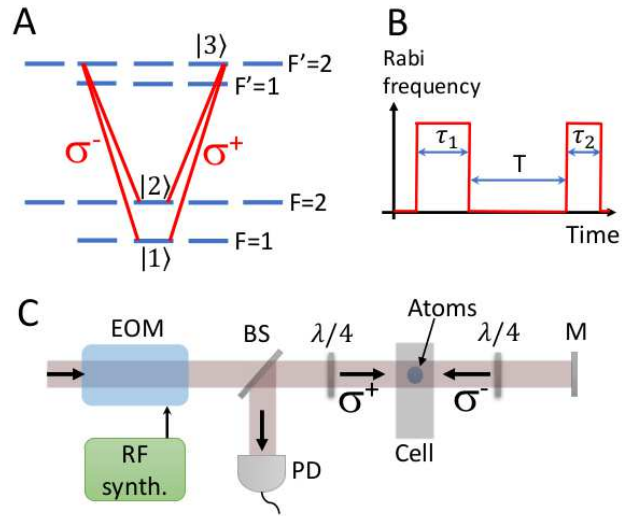}
    \caption{
    \label{Figure1_scheme_exp_setup}
    A: Energy-level scheme of $^{87}$Rb D$_1$ transition and the $\sigma ^+-\sigma^-$ CPT configuration. The levels associated with the three-level model are denoted $|1\rangle$, $|2\rangle$ and $|3\rangle$. B: The timing sequence of Ramsey-CPT interrogation. A first CPT pulse of duration $\tau_1$ is applied to the atoms (typically up to $1$ ms), driving them to a coherent superposition of two levels (dark-state). The atoms then evolve in the dark for a duration of T (dark-period, typically a few milliseconds). After the dark-period, a short second CPT pulse is applied ($\sim 50$ $\mu$s), and its transmission effectively measures the phase difference between the atomic oscillator and the RF synthesizer. C: The experimental setup. Laser light is modulated by an electro-optic modulator (EOM) driven by a RF synthesizer to produce equal-intensity CPT bands. The light then impinges on a cloud of laser-cooled $^{87}$Rb atoms in free-fall (cooling apparatus not shown) with $\sigma ^+$ polarization, passes through a quarter wave-plate ($\lambda/4$), retro-reflected by a mirror (M) and passes again through the atoms with $\sigma ^-$ polarization. The mirror is positioned to match the phases of the dark-states created by the ongoing and reflected beams. A beam-splitter (BS) is used to measure the reflected beam on a photo-diode (PD).
    }
\end{figure}

We measure the dark-state pumping and decay rates by applying CPT light fields to a cloud of free-falling laser-cooled $^{87}$Rb atoms in the NIST cold-atom CPT clock apparatus (see details in \cite{BlanshanPRA2015,LiuPRAppl2017}). A magneto-optical trap (MOT) is applied for $20$ ms, followed by a $3$ ms optical molasses, to trap and cool about a million atoms to $10$ $\mu$K. The atoms are then allowed to free fall with a static quantization magnetic field of $4.4$ $\mu$T oriented parallel to the CPT beam propagation direction. During the free-fall, a CPT interaction is formed with the $\sigma ^+-\sigma ^-$ polarization configuration to the $F'=2$ level \cite{TaichenachevJETPL2004} to achieve high-contrast CPT resonances, as shown in Fig. \ref{Figure1_scheme_exp_setup}.A. The two light-field frequencies (with equal intensity) are obtained by passing the beam from a distributed Bragg reflector laser through an electro-optic modulator (EOM) driven at about $6.835$ GHz which is the $^{87}$Rb ground-state hyperfine splitting. The EOM is driven by a RF synthesizer, and the carrier and the $(-1)$-order side-band are used as the two CPT frequencies. We used both standard Rabi CPT spectroscopy and Ramsey-CPT spectroscopy \cite{ZanonPRL2005} in this study. The Rabi CPT spectroscopy is a measurement of the steady-state transmission of the CPT beam reached at the end of a single pulse vs. the frequency difference between the CPT fields (the RF frequency driving the EOM). The Ramsey-CPT method is applied by sending two CPT laser pulses of duration $\tau_1,\tau_2$ separated by a dark-period of duration T and measuring the transmission of the short second pulse, as shown in Fig. \ref{Figure1_scheme_exp_setup}.B. The experimental setup we use to apply the $\sigma ^+-\sigma ^-$ configuration and measure the transmission of the second CPT pulse is shown in Fig. \ref{Figure1_scheme_exp_setup}.C.\\
The transmission of the second Ramsey-CPT pulse reflects the phase difference accumulated between the RF synthesizer and the atomic oscillator which in this case is a coherent dark state with a phase oscillating at the ground-state hyperfine splitting frequency. The Ramsey fringe pattern results from scanning this phase difference by scanning the RF synthesizer's frequency while measuring the second pulse transmission. The amplitude of the Ramsey fringes is proportional to the fraction of the atoms that occupy the dark state. By tracing the Ramsey fringe amplitude under different experimental conditions, the dynamics of pumping atoms into the dark state, as well as dark-state decay can be monitored. In section \ref{Measurement of dark-state pumping rate}, a measurement of the pumping into the dark-state is presented and its influence on the light shifts in Ramsey spectroscopy is presented in section \ref{The effect of the pumping rate on light-shifts}. The decay rate of the dark state is studied in section \ref{Measurement of dark-state decay}.\\

\section{Measurement of the dark-state pumping rate}
\label{Measurement of dark-state pumping rate}
\begin{figure}[t]
    \includegraphics[width=8.6cm]{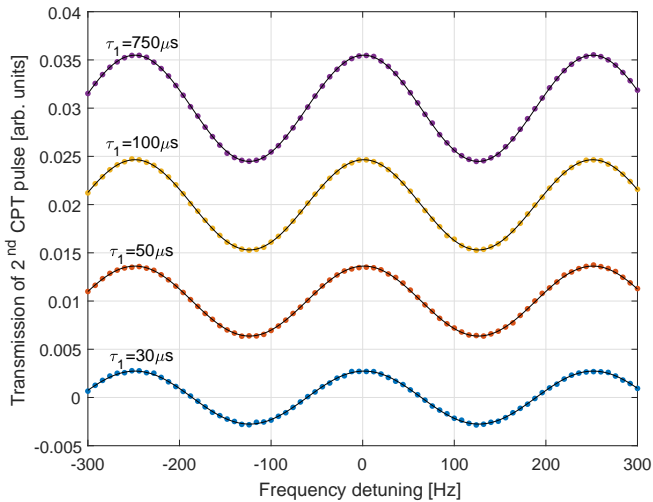}
    \caption{
    \label{Figure2_Ramsey_fringes_different_tau1s}
    Measured Ramsey fringes (dots) for different CPT preparation pulse duration ($\tau_1$) with CPT beam intensity of $1.33$ W/$m^2$ and fits to a sinusoidal function (thin solid lines). The curves are vertically shifted for clarity. For each preparation pulse duration, a set of Ramsey measurements was performed versus the frequency detuning between the two CPT fields, resulting in a Ramsey fringes pattern in the second CPT pulse transmission. The amplitude of the Ramsey fringes is proportional to the fraction of the atoms that are in the dark-state when the second CPT pulse is applied. As the preparation pulse duration is increased in the range $30-750$ $\mu$s, the amplitude of the Ramsey fringes increases due to the pumping of a larger fraction of the atoms into the dark-state.
    }
\end{figure}

\begin{figure}[t]
    \includegraphics[width=8cm]{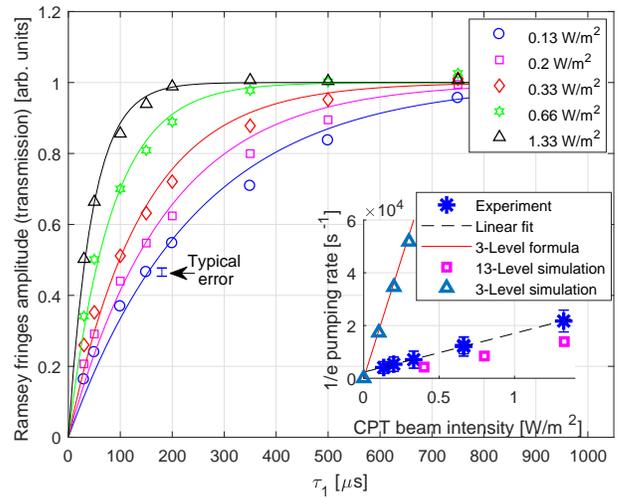}
    \caption{
    \label{Figure3_pumping_rate_vs_intensity}
    Ramsey fringe amplitude vs. pumping time for different CPT beam intensities. The symbols show the measured Ramsey fringes amplitude (normalized for clarity) vs. the first CPT pulse duration $\tau_1$ for different CPT laser beam intensities. The solid curves are a fit to an exponential function. As expected, at higher intensities the pumping rate into the dark state is higher. The inset shows the measured exponential pumping rate (asterisks) vs. the CPT beam intensity along with the expected rate from the three-level analytic formula (Eq. \eqref{dark_state_pumping_rate}, solid line). The measured pumping rates depend linearly on the intensity, however there is about a factor of ten discrepancy from pumping rate expected from the three-level formula. Also shown in the inset are the prediction of a 3-level numerical model (triangles) and a 13-level numerical model (squares), showing that most of the discrepancy can be explained by the multi-level Zeeman structure of the atom. 
    }
\end{figure}

\begin{figure}[t]
    \includegraphics[width=8cm]{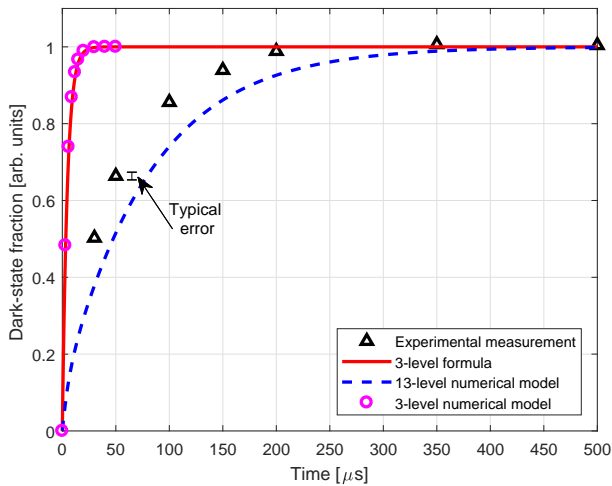}
    \caption{
    \label{Figure4_dark_state_pumping_exp_13L_model}
    The measured dark-state pumping curve for a CPT beam intensity of $1.33$ W/$m^2$ and a comparison to various models. The experimental pumping rate (triangles) is a factor of ten slower than predicted by the 3-level analytic formula (Eq. \eqref{dark_state_pumping_rate}, solid line). In the 13-level numerical model (dashed line) the dark-state fraction is obtained by finding the absolute value of the density-matrix coherence term between the two levels of the clock transition ($|F=1;m_F=0\rangle$ and $|F=2;m_F=0\rangle$). The 13-level numerical model shows a reasonable agreement with the measurements using no fit parameters. A 3-level numerical model (circles) reproduces the results of the analytic formula.
    }
\end{figure}

When a CPT light field is applied to an atom in an incoherent quantum state, the atom is optically pumped into a coherent dark state. The atom is driven into an excited electronic configuration by the CPT field and subsequently spontaneously decays into the ground state. Since the dark state does not couple to the CPT field and other states do, the atomic population is eventually driven into the dark state. The rate of this process depends on the properties of the CPT field and on the properties of the atom, including the internal electronic structure, the optical coupling strength and the decay rates within the atom. This process of pumping into the dark state is usually described by a single exponential pumping rate $\Gamma_p$, and for a three-level system the pumping rate is given by \cite{HemmerJOSAB1989,GuerandelIEEE2007,PollockPRA2018}:
\begin{equation}\label{dark_state_pumping_rate}
    \Gamma_p = \frac{1}{4}\frac{|\Omega_{1}|^2+|\Omega_{2}|^2}{\gamma_\text{opt}}.
\end{equation}
The CPT-field Rabi frequencies are given by $\Omega_{1}=d_{31}E_{1}/\hbar$ and $\Omega_{2}=d_{32}E_{2}/\hbar$ for the transitions $|1\rangle\leftrightarrow|3\rangle$ and $|2\rangle\leftrightarrow|3\rangle$, respectively ($d_{31}$ and $d_{32}$ are the reduced matrix elements of the dipole moments for these transitions) and $\gamma_\text{opt}$ is the rate of decoherence of the optical transitions $|1\rangle\leftrightarrow|3\rangle$ and $|2\rangle \leftrightarrow|3\rangle$ ($\gamma_\text{opt}=\gamma/2$ for pure spontaneous relaxation where $\gamma$ is the upper level decay rate).\\
A more realistic and complex structure of the atoms, as well as more complex polarization configurations ($\sigma ^+-\sigma^-$, lin $||$ lin) might affect the pumping rate into the dark state and in-turn influence other properties of the CPT resonances including their light shift. For this purpose, a quantitative measurement of the pumping time into the dark state was performed by tracking the amplitude of the Ramsey fringes for different durations of the first CPT pulse in the Ramsey cycle ($\tau_1$). Figure \ref{Figure2_Ramsey_fringes_different_tau1s} depicts a set of measured Ramsey fringes with $\tau_1$ in the range $30-750$ $\mu$s (curves shifted vertically for clarity). All other experimental parameters, including $T=4$ ms and $\tau_2=50$ $\mu$s, were kept constant. As $\tau_1$ increases in this range, the amplitude of the Ramsey fringes grows. This is a signature of the pumping process of the atoms into the dark-state which we use here to measure the dark state pumping rate. \\
Figure \ref{Figure3_pumping_rate_vs_intensity} presents normalized Ramsey fringe amplitudes as a function of the pumping pulse duration for different CPT laser total intensities. For each intensity, the measurements show the gradual pumping into the dark state. By fitting an exponential curve to the measurements (solid lines) the 1/e pumping rate for each intensity is extracted. The inset of Fig. \ref{Figure3_pumping_rate_vs_intensity} shows the measured 1/e pumping rate into the dark-state vs. the CPT beam intensity (asterisks), as well as the predicted pumping rate from the three-level analytic formula (solid line). The three-level formula takes into account the three levels denoted by $|1\rangle$, $|2\rangle$ and $|3\rangle$ in Fig. \ref{Figure1_scheme_exp_setup}.A and assumes only $\sigma^+$ radiation. As expected from Eq. \eqref{dark_state_pumping_rate}, the measured pumping rates exhibit a linear dependence on the intensity (dashed-line). However, the measured pumping rate is about one order of magnitude smaller than predicted by three-level model that leads to Eq. \eqref{dark_state_pumping_rate}. We attribute this discrepancy to the full Zeeman structure of the atom combined with the complex scheme involved in a $\sigma ^+-\sigma^-$ polarization interrogation. We note that using the measured pumping rates presented here, we observed an improved agreement between the theory for off-resonant light shifts and experimental measurements \cite{PollockPRA2018}. In \cite{ZanonPRL2005}, measurements of the dark-state pumping rate for Cesium vapor in cells with buffer gas are presented using the lin $\perp$ lin configuration, also showing a slower rate compared to the formula in Eq. \eqref{dark_state_pumping_rate} \cite{PrivComm}.\\
To verify that the multi-level Zeeman structure is the main cause for the discrepancy from the analytic formula, we solve a 13-level numerical model and simulate the pumping process into the dark-state. The model contains all the Zeeman manifold of levels F=1, F=2, F'=2 of the $^{87}$Rb D$_1$ transition (13 levels). We apply resonant light fields with $\sigma^+$ and $\sigma^-$ polarizations for both $F=1\rightarrow F'=2$ and $F=2\rightarrow F'=2$ transitions, taking into account the Zeeman shift within the ground-state. We use the density matrix formalism and numerically solve its dynamic evolution, introducing decay using Lindblad operators. Due to the cooling process, the initial conditions assume that all the Zeeman sub-levels of the F=2 level are equally populated without any coherence. 
Figure \ref{Figure4_dark_state_pumping_exp_13L_model} shows the dark-state pumping curve calculated by this numerical model (dashed line), as compared with our experimental measurements (triangles) and the rate predicted by Eq. \eqref{dark_state_pumping_rate} (solid line). The 13-level model agrees with the experimental pumping rate within $\sim30\%$. We also used the same numerical model to simulate a 3-level model and found that it agrees well with Eq. \eqref{dark_state_pumping_rate} (circles in Fig. \ref{Figure4_dark_state_pumping_exp_13L_model}).\\
The inset of Fig. \ref{Figure3_pumping_rate_vs_intensity} shows the dark-state pumping rate vs. the CPT beam intensity for both the 3-level (triangles) and the 13-level (squares) numerical models. While the 3-level model agrees well with the 3-level analytic formula, the 13-level model shows a reasonable agreement with the experimental results (within $\sim30\%$). We attribute the part of the discrepancy to the fact that the pumping process is not purely exponential, as can be see by the deviation form the exponential fits in Fig. \ref{Figure3_pumping_rate_vs_intensity}.

\section{The effect of the pumping rate on light-shifts}
\label{The effect of the pumping rate on light-shifts}
\begin{figure}[t]
    \includegraphics[width=8cm]{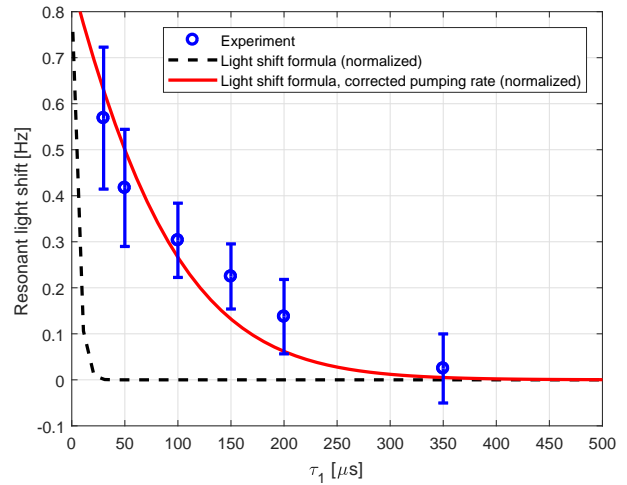}
    \caption{
    \label{Figure5_light_shift_vs_tau1}
    The measured light-shift of the central fringe (circles) using Ramsey-CPT spectroscopy with $T=4$ ms vs. the duration of the first CPT pulse $\tau_1$. The theoretical formula for the light shift \cite{HemmerJOSAB1989} (dashed line) predicts a faster reduction in the light shift with the increase in the pumping pulse duration compared wit the experiment. By correcting the theoretical formula using the pumping rate deduced from the 13-level model (solid line) a good fit to the measurement is obtained.
    }
\end{figure}

The slower pumping rate into the dark-state might influence measurements utilizing these unique quantum states. For example, in an atomic clock utilizing Ramsey-CPT spectroscopy \cite{HemmerJOSAB1989,ZanonPRL2005} the light-shift is related to the rate at which the atomic population is pumped into the dark-state \cite{HemmerJOSAB1989}. In \cite{HemmerJOSAB1989,ShahriarPRA1997} a three-level theory is presented, predicting the light shift in the central fringe of a Ramsey-CPT spectroscopy. Using the notation of \cite{ShahriarPRA1997} the angular phase-shift of the central Ramsey-CPT fringe $\phi$ (where $\phi/T$ is the frequency shift) is given by:
\begin{equation}\label{resonant_light_shift_formula}
    \tan(\phi)\propto\frac{e^{-\alpha\tau_1}}{1-e^{-\alpha\tau_1}}\sin(\beta\tau_1),
\end{equation}
where
\begin{equation}\label{alpha_beta_definition}
    \alpha=\frac{1}{2}\frac{\Omega^2\gamma}{\gamma^2+4\delta^2} ; \beta=\frac{1}{2}\frac{\Omega^2\delta}{\gamma^2+4\delta^2},
\end{equation}
$\Omega^2=|\Omega_{1}|^2+|\Omega_{2}|^2$ and $\delta$ is the one photon detuning.
We used the Ramsey-CPT spectra measurements presented in Fig. \ref{Figure2_Ramsey_fringes_different_tau1s} to find the light-shift of the central Ramsey-CPT fringe for different pumping durations. We note that we previously measured this shift in an alternative method, by locking an atomic clock on the central fringe and measuring the shift of the clock \cite{LiuPRAppl2017}, and the two measurements are consistent. Figure \ref{Figure5_light_shift_vs_tau1} shows the results of this measurement (circles), demonstrating the drop in the light shift when the pumping duration increases, as expected by theory \cite{HemmerJOSAB1989,ShahriarPRA1997} and observed previously \cite{LiuPRAppl2017}. The theoretical formula (Eq. \eqref{resonant_light_shift_formula}) deviates substantially from the experimental results (dashed line in Fig. \ref{Figure5_light_shift_vs_tau1}). However, by replacing the pumping rate $\Omega^2$ by the rate calculated from the 13-level numerical model described above, a good fit to the experimental data is obtained (solid line in Fig. \ref{Figure5_light_shift_vs_tau1})

\section{Measurement of dark-state decay}
\label{Measurement of dark-state decay}
\begin{figure}[t]
    \includegraphics[width=8.2cm]{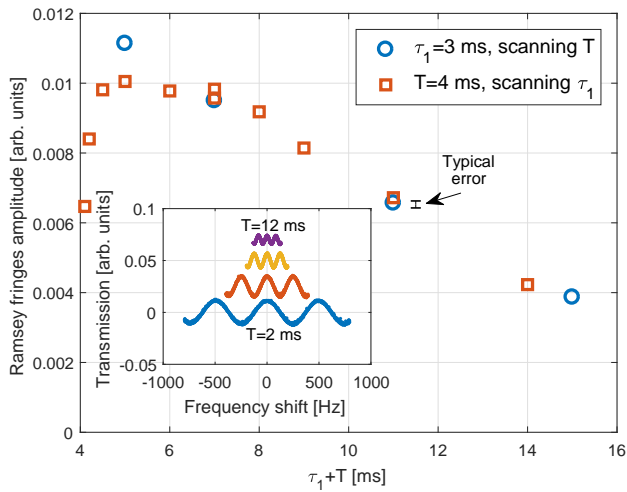}
    \caption{
    \label{Figure6_Ramsey_fringes_decay}
    Decay of Ramsey-fringe amplitude at long pumping durations and dark periods. The graph shows the measured Ramsey fringes amplitude vs. the total time from the beginning of the first CPT pulse to the beginning of the second CPT pulse ($T+\tau_1$). In one measurement (squares) the dark-period was kept constant ($T=4$ ms) and the CPT pumping pulse duration was scanned in the range $\tau_1=0.1-10$ ms. In the second measurement (circles) the CPT pumping pulse duration was kept constant ($\tau_1=3$ ms) and the dark period was scanned in the range $T=2-12$ ms (the inset shows the measured Ramsey fringes for this case with $T=2,4,8,12$ ms). It is evident that the decay rate depends dominantly on the total time from the MOT shutoff (and not on the time from the end of the CPT pumping pulse). This fact suggests that the dominant decay mechanism is the mechnical motion of the atoms out of the probing region (rather than decay of their quantum state).
    }
\end{figure}

\begin{figure}[t]
    \includegraphics[width=8.6cm]{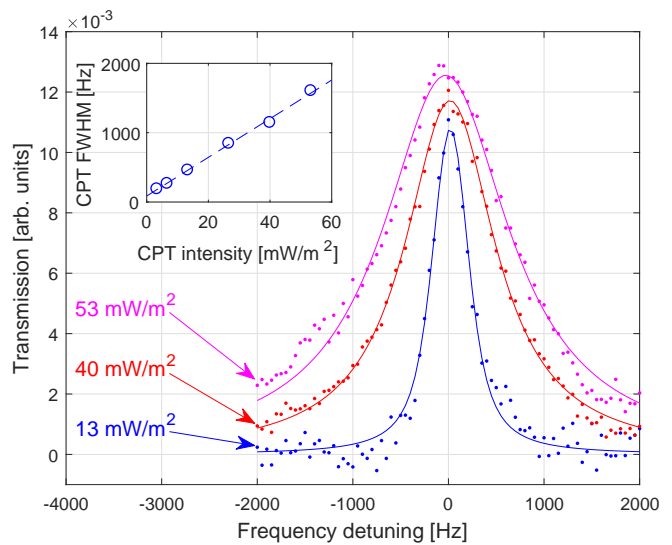}
    \caption{
    \label{Figure7_CPT_power_broadening}
    The CPT resonance line-shape at different CPT intensities. The transmission of the first Ramsey pulse (at times longer than the CPT pumping time) is measured while scanning the frequency difference of the two CPT fields (dots). By fitting the experimental data to a Voigt profile (solid lines) the line-width of the CPT resonance is measured (as expected the line-shape is nearly a pure Lorentzian). The inset shows the measured full-width at half-maximum (FWHM) vs. the CPT intensity. As expected, a linear power broadening curve is observed, and the y-axis intercept is the power-broadening-free FWHM.
    }
\end{figure}

The Ramsey fringe amplitude is a measure of the dark-state population in the atomic ensemble (see section \ref{Measurement of dark-state pumping rate}) and can also be used to monitor the decay of the dark state. The decay of the dark-state in the atomic ensemble can originate from internal processes in an atom (natural decoherence, collisions) as well as from external processes (dephasing between the dark states of different atoms in the ensemble, motion of the atoms out of the probing region). Figure \ref{Figure6_Ramsey_fringes_decay} depicts a measurement of the Ramsey fringe amplitude vs. the total time from the beginning of the first CPT pulse to the beginning of the second CPT pulse, $\tau_1+T$. In one measurement (squares) we keep the dark-period constant at $T=4$ ms and increase the pumping duration $\tau_1$ up to $10$ ms. After the initial increase in the Ramsey fringe amplitude (examined in details in section \ref{Measurement of dark-state pumping rate}), the amplitude decreases as the pumping time is further increased, demonstrating the ensemble's dark-state decay. In another measurement (circles) we keep the pump duration constant at $\tau_1=3$ ms and scan the dark period in the range $T=2$ to $12$ ms, showing a decrease in the fringe amplitude as the dark-period increases (the inset of figure \ref{Figure6_Ramsey_fringes_decay} shows the measured Ramsey fringes for $T=2,4,8,12$ ms). In both measurements the decay of the dark-state in the atomic ensemble is apparent. However, the decoherence process starts to take effect only after the first CPT pulse is over (during the pumping pulse any decoherence is immediately compensated by the faster dark-state pumping process). The similarity of the decay rates in the two measurements suggests that the decoherence is negligible. The universal dependence of the amplitude on the total time since the cooling light shut off ($\tau_1+T$) suggests that the motion of the atoms out of the CPT beam is the dominant decay mechanism (since that depends only on the time since cooling light shut off, and not on the timing of the pumping CPT pulse). \\
In order to verify that the decoherence of the dark-state is indeed negligible, we have measured the CPT resonance line-width at various experimental conditions, in order to calculate the effective natural line-width of the CPT resonance. Figure \ref{Figure7_CPT_power_broadening} depicts the measured CPT resonances at three different intensities of the CPT beam (dots) fitted by a Voigt profile (solid lines). The inset of Fig. \ref{Figure7_CPT_power_broadening} shows the measured power-broadening of the CPT resonance, which shows a linear dependence of the CPT resonance full-width at half-maximum (FWHM) on the CPT beam intensity. The y-axis intercept of this power broadening curve is the power-broadening-free width. The measured width of the CPT resonance is also broadened due to the finite duration of the CPT pulse (Fourier broadening). By repeating this measurement for different CPT pulse durations (in the range from $\tau_1=1-10$ ms), the Fourier broadening was also subtracted. Finally we found that the natural CPT line-width (power-broadening free and Fourier-broadening free) was $6\pm20$ Hz. This result verifies the long coherence time of the atomic dark-state.
\section{Discussion}
\label{Discussion}
In conclusion, we have measured the pumping rate and the decay rate of a dark-state in an ensemble of cold $^{87}$Rb atoms using a cold-atom CPT appartus. We found that the pumping process into the dark-state is about ten times slower than expected by the commonly used three-level analytic formula (Eq. \eqref{dark_state_pumping_rate}), which can dramatically affect systematic evaluations of the light shift in precision measurements. We attribute this discrepancy to the more complex interaction of a multi-level atom in the $\sigma^+-\sigma^-$ polarization configuration compared with a three-level formula. We demostrated that a 13-level numerical model shows better agreement with the experimental results. When using the pumping rates calculated by the 13-level numerical model in the Ramsey-CPT light-shift formula \cite{HemmerJOSAB1989,ShahriarPRA1997} we obtain a good agreement with experimental measurements of the shift of the central Ramsey-CPT fringe. We also measured the decay of the dark-state in the atomic ensemble and found that in our setup it is dominated by the mechanical motion of the atoms out of the probing region. The natural CPT line-width, without power-broadening and Fourier-brodening, was measured to be $6\pm20$ Hz, demonstrating the long coherence time of dark states in free-falling cold atoms.
\begin{acknowledgments}
The authors acknowledge R. Boudot, Y.-J. Chen and K. Beloy for review of the manuscript and helpful discussions. V. I. Yudin was supported by the Russian Science Foundation (Project No. 16-12-10147). This work is a contribution of NIST, an agency of the U.S. government, and is not subject to copyright. 
\end{acknowledgments}
\bibliography{bib_CACPT}
\bibliographystyle{apsrev4-1}
\end{document}